\documentclass[12pt,notitlepage,final,letterpaper]{iopart}
\usepackage{iopams}
\usepackage{times,txfonts}
\usepackage{hyperref,graphicx,subfigure,amsbsy,bbm,footmisc,color}

\newcommand{\prl}{{\it Phys. Rev. Lett.} }
\newcommand{\rmp}{{\it Rev. Mod. Phys.} }
\newcommand{\pra}{{\it Phys. Rev.} A }
\newcommand{\pla}{{\it Phys. Lett.} A }
\newcommand{\prd}{{\it Phys. Rev.} A }
\newcommand{\njp}{{\it New J. Phys.} }
\newcommand{\arx}{{\it e--print} }
\newcommand{\R}{\mathbbm{R}}

\newcommand{\sy}[1]{{\rm Sp}_{(#1,\R)}}
\newcommand{\id}{\mathbbm{I}}
\newcommand{\text}[1]{{\rm #1}}
\newcommand{\hh}{\mathcal{H}}
\newcommand{\T}{^{\sf T}}
\newcommand{\gr}[1]{\boldsymbol{#1}}

\newcommand{\be}{\begin{equation}}
\newcommand{\ee}{\end{equation}}
\newcommand{\bea}{\begin{eqnarray}}
\newcommand{\eea}{\end{eqnarray}}
\newcommand{\eq}[1]{Eq.~(\ref{#1})}

\newcommand{\ket}[1]{|#1\rangle}
\newcommand{\bra}[1]{\langle#1|}

\newcommand{\op}[1]{\hat{#1}}
\newcommand{\adj}[1]{{#1}^{\dag}}
\newcommand{\comm}[2]{\left[#1,#2\right]}
\newcommand{\sig}{{\gr\sigma}}
\newcommand{\gam}{\boldsymbol{\gamma}}

\begin{document}

\title[Continuous variable methods in relativistic quantum information]{Continuous variable methods in relativistic quantum information: Characterisation of quantum and classical correlations of scalar field modes in noninertial frames}

\author{Gerardo Adesso, Sammy Ragy, and Davide Girolami}
\address{School of Mathematical Sciences, The University of Nottingham, University Park, Nottingham NG7 2RD, United Kingdom}
\ead{gerardo.adesso@nottingham.ac.uk}

\begin{abstract}
We review a recently introduced unified approach to the analytical quantification of correlations in Gaussian states of bosonic scalar fields by means of R\'enyi-$2$ entropy. This allows us to obtain handy formulae for classical, quantum, total correlations, as well as bipartite and multipartite entanglement. We apply our techniques to the study of correlations between two modes of a scalar field as described by observers in different states of motion. When one or both observers are in uniform acceleration, the quantum and classical correlations are degraded differently by the Unruh effect, depending on which mode is detected. Residual quantum correlations, in the form of quantum discord without entanglement, may survive in the limit of an infinitely accelerated observer Rob, provided they are revealed in a measurement performed by the inertial Alice.
\end{abstract}

\tableofcontents
\markboth{Continuous variable methods in relativistic quantum information}{}
\submitto{\CQG}

\section{Introduction}

Relativistic quantum information (RQI)
is a blooming area of research devoted to the study of quantum information concepts and processes under relativistic conditions \cite{peresreview,ternoreview,ivettenotes,eduthesis}.
Traditional streams of investigation in the domain of RQI have included the characterisation of entropy and entanglement between modes of a quantum field as perceived by observers in different states of motion \cite{peresreview,peresterno,alicefalls,alicefermion,unruhsharing,eduunruh,eduunruh2,bruschi,amplif}, the production of entangled particles in curved spacetimes and models of the expanding universe \cite{ball,moradi,eduparticles}, the investigation and applications of spacelike and timelike entanglement extracted from the quantum vacuum \cite{retzker,ralphtim}, the modification and generation of entanglement by moving cavities \cite{downes,alpha,motiongenerates,niconew}, and the analysis of quantum communication protocols such as teleportation and key distribution in noninertial reference frames \cite{alsing,bradler,nathan}.
The main theoretical ingredients for RQI ventures are a marriage of quantum field theory on one hand \cite{birelli}, and the formalism of quantum information theory \cite{nielsenchuang} on the other. For fermionic (Grassman) fields, described by field operators $\hat c$ subject to anticommutation relations $\{\hat c, \hat c^\dagger\}=1$,  one can investigate state properties and mode correlations by employing the quantum information techniques usually adopted for states of multi-qubit systems, where by `qubit' we mean a two-level quantum system \cite{nielsenchuang}. For bosonic scalar fields, described by field operators $\hat b$ satisfying canonical commutation relations $[\hat b, \hat b^\dagger]=1$, each mode lives in an infinite-dimensional Hilbert space and represents a so-called `continuous variable' system. Techniques from continuous variable quantum information  \cite{eisertplenio,brareview,ourreview,book,pirlandolareview} are thus potentially very useful for RQI investigations involving scalar fields.

In this paper we shall present a collection of relevant methods and measures to quantify state properties and correlations in modes of a free scalar field. We shall focus on Gaussian states and transformations \cite{ourreview}, as they arise naturally in a number of contexts in RQI \cite{unruhsharing,nathan,birelli,ahnkim,massar,buchi,mresonance} (see Ref.~\cite{niconew} for a recent overview), and they enjoy tractable mathematical expressions. As Gaussian states are the states of {\it any} physical system in the harmonic approximation \cite{klauder,schuch}, or so-called `small oscillations' limit, they lend themselves as first-choice testbeds for novel theoretical investigations; therefore it is no surprise that their role in  RQI ventures has become so prominent. Moreover, some transformations, such as those associated to the change of coordinates between Minkowski and Rindler observers in flat spacetimes, are naturally associated---with no approximation---to Gaussian operations \cite{birelli}. In fact, the Unruh effect on scalar fields \cite{unruh}, and the closely related Hawking effect in the presence of a black hole \cite{hawking}, can be formally described in terms of the action of a Gaussian amplification channel \cite{bradler,mariona}. Gaussian states are furthermore particularly easy to prepare and control in a range of setups including primarily quantum optics, atomic ensembles, trapped ions, optomechanics, as well as hybrid interfaced networks thereof \cite{book}. This could make them candidates of choice for the implementation of explorative experiments to, at least,  simulate relativistic phenomena in the quantum optical setting, e.g., in the spirit of Ref.~\cite{faccio} (keeping in mind the warnings advanced in Ref.~\cite{com}).

It is however appropriate to stress  that the above mentioned {\it liaison} between relativity and quantum information holds, to date, to a formal equivalence at mathematical level.  We stress that the main purpose of our work is in fact to provide mathematical tools that might be useful for ongoing and future research in RQI: In this respect, our work aims to deploy solid theoretical methods and results rather than to develop actually feasible experimental proposals. The reader must be aware that concerns about the ultimate physical meaningfulness of certain RQI findings have been advanced. For instance, global relativistic quantum field modes cannot be measured by ideal measurements, i.e., measurements that map eigenstates  of an observable into themselves  \cite{fay1}.  Still, analysing how basic quantum field theory predictions affect fundamental quantum correlations between global field modes may be instructive to better grasp the basics of the mechanisms involved. Surely, translating the results obtained in this scenario to carry out real experiments is not straightforward at all, and more refined approaches should thus be adopted. Here we limit ourselves to mention some recent proposals regarding localised projective measurements \cite{edu1} and particle detector models \cite{edu2}, deferring the discussion to the concluding section for additional remarks.

Let us briefly recall some previous works related to our analysis. A comprehensive characterisation of the degradation and redistribution of entanglement between modes of a bosonic scalar field was developed in \cite{unruhsharing} by means of Gaussian quantum information techniques. In that paper, entanglement and total correlations in the state of two  field modes---described as a two-mode squeezed (Gaussian) entangled state from a fully inertial perspective---were found to degrade if one or both observers undergo uniform acceleration (see also \cite{ahnkim}). In the case of one inertial (Alice) and one noninertial observer (Rob, living in Rindler region $I$), the lost entanglement was interpreted as redistributed genuine tripartite entanglement among Alice, Rob, and an observer (known in the literature as anti-Rob) living in the causally disconnected Rindler region $II$. A similar analysis for fermionic fields was reported in \cite{alicefermion}.

Entanglement \cite{entanglementreview} is, however, not the only form of quantum correlation. A finer description of quantumness versus classicality of correlations in bipartite quantum states has been recently put forward \cite{zurek,vedral}. Measures such as the one-way classical correlation and the quantum discord have been now computed \cite{giordaparis,adessodatta,mistagauss,renyi} and measured experimentally \cite{expdiscordgauss} for Gaussian states, and provide a deeper insight into the nature of correlations compared to the entanglement/separability dichotomy. In rough terms, classical correlations correspond to how much, at most, the ignorance that one observer (say Alice) has about the marginal state of her subsystem, is reduced when the other observer (say Rob) performs a measurement on his subsystem \cite{vedral}. This has to be maximised over all possible measurements on Rob's side. Complementarily, the genuinely quantum correlations, as captured by the quantum discord \cite{zurek}, are those destroyed in the above described process of a marginal measurement on one subsystem only. Including the optimisation over measurements, this corresponds to how much, at least, a marginal measurement disturbs the state of a composite system, which is a distinctively quantum feature. In this sense general quantum correlations are always {\it revealed} by means of  marginal  measurement processes \cite{pianiadesso}. Formal definitions of these quantities will be provided later; the interested reader can refer e.g.~to a recent review \cite{modireview} for further details. It is immediately clear that the above concepts for classical and quantum correlations have, unlike entanglement, an intrinsically non-symmetric nature. If we swap over the roles of the two observers, quite different results can be obtained. In particular, it is possible that quantumness of correlations can be revealed, or detected, by measurements on one subsystem, but not on the other. This is precisely what will be found to happen in the state of two scalar field modes in the limit of infinite acceleration of Rob: Quantum discord is destroyed (like entanglement)---and classical correlations unaffected---if Rob is the measuring party, while enduring quantum correlations remain detectable if the inertial observer Alice is in charge of the measurement (see also \cite{dattaunruh}).

The paper contents and structure are as follows.
In Sec.~\ref{secgauss} we  review the formalism of continuous variable Gaussian states and their informational properties \cite{ourreview}; we  adopt a recently introduced unified approach to the study of Gaussian correlations (including entanglement, classical, quantum, and total correlations) by means of R\'enyi-$2$ entropy \cite{renyi}.  In Sec.~\ref{secalice} we  recall the basics of the Unruh effect for scalar fields, and we apply the introduced techniques to characterise how various forms of correlations are affected by acceleration of one or both observers detecting two  field modes which are in an entangled Gaussian state from a fully inertial perspective. In Sec.~\ref{secconcl} we draw our concluding remarks and outline relevant perspectives.

\section{Gaussian states, operations, information and correlation measures}\label{secgauss}

\subsection{Continuous variable systems}
A continuous variable system of $N$ canonical bosonic modes
is described by a Hilbert space $\hh=\bigotimes_{k=1}^{N} \hh_{k}$
resulting from the tensor product structure of infinite-dimensional
Fock spaces $\hh_{k}$'s, each of them associated to a single mode \cite{eisertplenio,brareview,ourreview}.
For instance, one can think of a non interacting quantised
scalar field (such as the electromagnetic field), whose Hamiltonian
\begin{equation}\label{CV:Ham}
\op{H} = \sum_{k=1}^N \hbar \omega_k \left(\adj{\op{b}}_k\op{b}_k +
\frac12\right)\,,
\end{equation}
describes a system
of an arbitrary number $N$ of harmonic oscillators of different frequencies,
the {\em modes} of the field.
Here $\op{b}_k$ and $\adj{\op{b}}_k$ are the  annihilation and
creation operators of an excitation in mode $k$ (with frequency
$\omega_k$), which satisfy the bosonic commutation relation
\begin{equation}\label{CV:comm}
\comm{\op{b}_k}{\adj{\op{b}}_{k'}}=\delta_{kk'}\,,\quad
\comm{\op{b}_k}{\op{b}_{k'}}=\comm{\adj{\op{b}}_k}{\adj{\op{b}}_{k'}}=0\,.
\end{equation}
From now on we shall assume for convenience natural units with
$\hbar=c=1$. The corresponding quadrature phase operators (`position'
and `momentum')  for each mode are defined as
\[  \hat q_{k} = \frac{(\op b_{k}+\op b^{\dag}_{k})}{\sqrt{2}}\,, \quad
  \hat p_{k} = \frac{(\op b_{k}-\op b^{\dag}_{k})}{i \sqrt 2}\,.
\]
We can group together the canonical operators in the vector
\be\label{CV:R}
\boldsymbol{\hat R}=(\hat{q}_1,\hat{p}_1,\ldots,\hat{q}_N,\hat{p}_N)\T
\in \mathbbm{R}^{2N}\,,
\ee
which enables us to write in compact form the  bosonic commutation
relations between the quadrature phase operators, \be
[\hat{R}_k,\hat{R}_l]= i\Omega_{kl} \; ,\label{ccr}\ee where
$\gr\Omega$ is the $N$-mode symplectic form \be
\gr\Omega=\bigoplus_{k=1}^{N}\gr\omega\, , \quad \gr\omega=
\left(\begin{array}{cc}
0&1\\
-1&0
\end{array}\right)\, . \label{symform}
\ee

The space $\hh_k$ is spanned by the Fock basis $\{\ket{n}_k\}$ of
eigenstates of the number operator $\hat{n}_k = \hat b_k^{\dag}\hat b_k$, representing the Hamiltonian of the noninteracting mode via
\eq{CV:Ham}. The Hamiltonian of each mode is bounded from below, thus
ensuring the stability of the system. For each mode  $k$
there exists a different vacuum state $\ket{0}_k\in \hh_k$
such that $\hat b_k\ket{0}_k=0$.
The vacuum state of the global Hilbert space will be denoted by
$\ket{0}=\bigotimes_k \ket{0}_k$.

The states of a continuous variable system are the set of positive trace-class
operators $\{\rho\}$ on the Hilbert space $\hh=\bigotimes_{k=1}^N
\hh_k$. Alternatively, for continuous variable systems, any state can be conveniently  described by the so-called Wigner quasi-probability distribution, obtained as the Wigner-Weyl transform from $\rho$ \cite{barnett}, and defined as
\begin{equation}\label{wigfunc}
W_\rho(\boldsymbol{\xi})=\frac{1}{\pi^N} \int_{{\mathbbm R}^{2N}}
\chi_\rho(\boldsymbol{\kappa}) \,{\rm e}^{i\boldsymbol{\kappa}^{\sf T} \gr{\Omega} \boldsymbol{\xi}}\,{\rm d}^{2N}\boldsymbol{\kappa}\,,
\end{equation}
where  $\boldsymbol{\xi}$ and $\boldsymbol{\kappa}$ belong to the real $2N$-dimensional space $\Gamma=({\mathbbm R}^{2N},\gr\Omega)$, which is called {\em phase
space} in analogy with classical Hamiltonian dynamics, and $\chi_\rho$ is the characteristic function of $\rho$,
\begin{equation}\label{charfunc}
\chi_\rho(\boldsymbol{\kappa}) = \text{tr}\,[\rho \hat D(\boldsymbol{\kappa})]\,,
\end{equation}
with  \begin{equation}\label{weyl0}
 \hat{D}(\boldsymbol{\kappa}) = {\rm e}^{i\boldsymbol{\hat{R}}^{\sf T} \gr\Omega \boldsymbol{\kappa}}\end{equation}
being the Weyl displacement operator.

\subsection{Gaussian states}
The set of Gaussian states is, by definition, the set of states of a continuous variable system whose characteristic function and Wigner phase-space distribution are positive-everywhere, Gaussian-shaped functions.
Gaussian states, such as coherent, squeezed and thermal states, are thus completely specified by the first and second statistical moments of the phase quadrature operators. As the first moments can be adjusted by marginal displacements, which do not affect any informational property of the considered states, we shall assume them to be zero, $\langle \boldsymbol{\hat{R}}\rangle=0$ in all the considered states without loss of generality. The important object encoding all the relevant properties of a Gaussian state $\rho$ is therefore the covariance matrix (CM) $\gr\sigma$ of the second moments, whose elements are given by
\begin{equation}\label{sigmaij}
\sigma_{j,k}=\text{tr}[\rho \{\hat{R}_j,\hat{R}_k\}]\,.
\end{equation}
We can then write the Wigner distribution [Eq.~(\ref{wigfunc})] of a generic $N$-mode undisplaced Gaussian state in the compact form
\begin{equation}
\label{eq:wigner}
W_\rho(\boldsymbol{\xi}) = \frac{1}{\pi^N \sqrt{\det{\sig}}} \exp\big(-\boldsymbol{\xi}^{\sf T} \boldsymbol{\sig}^{-1} \boldsymbol{\xi}\big)\,.
\end{equation}

One can see
that in the phase space picture, the tensor product structure is replaced by a
direct sum structure, so that the $N$-mode phase space is $\Gamma =
\bigoplus_k \Gamma_k$, where $\Gamma_k=({\mathbbm R}^{2},\gr\omega)$ is
the marginal phase space associated with mode $k$.
Similarly, the CM for product states of the form $\otimes_k \rho_k$ will be the direct sum $\oplus_k \sig_k$ of individual covariance matrices for each subsystem. In particular, the global vacuum state $\ket{0}$ of a $N$-mode  scalar field is a Gaussian state with CM $\sig_0=\oplus_{k=1}^N \id$ where $\id$ denotes here the $2\times 2$ identity matrix.
If we partition our system into two subsystems $A$ and $B$ [$\hh=\hh_A \otimes \hh_B$], each grouping $N_A$ and $N_B$ modes respectively (with $N_A+N_B=N$), the CM of a $N$-mode bipartite Gaussian state $\rho_{AB}$ with respect to such a splitting can be written in the block form
\begin{equation}\label{eq:cms}
\sig_{AB} = \left(\begin{array}{c|c}
\sig_A & {\gr\varsigma}_{AB} \\ \hline
{\gr\varsigma}_{AB}^{\sf T} & \sig_B\,
\end{array}\right).
\end{equation}

We refer the reader to Refs.~\cite{eisertplenio,ourreview,pirlandolareview,schuch} for further details on the structural and formal description of Gaussian quantum states in phase space.

\subsection{Gaussian operations}
\paragraph{Gaussian unitaries.}
An important role in the theoretical and experimental manipulation
of Gaussian states is played by {\it unitary} operations $\hat{U}$ which preserve
the Gaussian character of the states on which they act. They are generated by Hamiltonian
terms which are at most quadratic in the field operators. By the  metaplectic
representation, any such unitary operation at the
Hilbert space level corresponds, in phase space, to a symplectic
transformation, that is, a linear transformation $\gr S$ which preserves
the symplectic form $\gr\Omega$: $\gr S\T\gr\Omega \gr S =
\gr\Omega$. Symplectic transformations on a $2N$-dimensional phase
space form the real symplectic group $\sy{2N}$. Such
transformations act linearly on first moments and by congruence on
covariance matrices, $\sig\mapsto \gr S \sig \gr S\T$.  Ideal beam splitters,
phase shifters and squeezers are all described by some kind of
symplectic transformation (see e.g.~\cite{francamentemeneinfischio}).
For instance, the two-mode squeezing operator
\begin{equation}\label{tmsU}
\hat{U}_{i,j}(r) = \exp[r (\hat b_{i}^\dagger \hat b_{j}^\dagger - \hat b_{i} \hat b_{j})]
\end{equation}
corresponds
to the symplectic transformation
\begin{equation}\label{tmsS}
\gr S_{i,j}(r)=\left(\begin{array}{cccc}
\cosh r&0&\sinh r&0\\
0&\cosh r&0&-\sinh r\\
\sinh r&0&\cosh r&0\\
0&-\sinh r&0&\cosh r
\end{array}\right)\, ,
\end{equation}
where the matrix is understood to act on the pair of modes $i$ and
$j$.

\paragraph{Gaussian measurements.}
In quantum mechanics, two main types of measurement processes are usually considered \cite{nielsenchuang}. The first type is constituted by projective (von Neumann) measurements, which are defined by a set of Hermitian positive operators $\{\Pi_i\}$ such that $\sum_i \Pi_i=\mathbbm{I}$ and $\Pi_i \Pi_j=\delta_{ij}\Pi_i$. A projective measurement maps a state $\rho$ into a state $\rho_i=\frac{\Pi_i \rho \Pi_i}{\text{tr}\{\Pi_i \rho \Pi_i\}}$ with probability $p_i={\text{tr}\{\Pi_i \rho \Pi_i\}}$. If we focus on a local projective measurement on the subsystem $B$ of a
bipartite state $\rho_{AB}$, say $\Pi_i=\mathbbm{I}_A\otimes \Pi_{iB}$, the subsystem $A$ is then mapped into  the conditional state $\rho_{A|\Pi_i}=\text{tr_B}\frac{\Pi_i \rho_{AB} \Pi_i}{\text{tr}\{\Pi_i \rho_{AB} \Pi_i\}}$.
The second type of quantum measurements are known as POVM (positive operator-valued measure) measurements and amount to a more general class compared to projective measurements. They are defined again in terms of a set of Hermitian positive operators $\{\Pi_i\}$ such that $\sum_i \Pi_i=\mathbbm{I}$, but they need not be orthogonal in this case. In the following, by `measurement' we will refer in general to a POVM.

 In the continuous variable case, the measurement operations mapping Gaussian states into Gaussian states are called Gaussian measurements. They can be realised experimentally by appending ancillae initialised in Gaussian states, implementing Gaussian unitary (symplectic) operations on the system and ancillary modes, and then measuring quadrature operators, which can be achieved e.g.~by means of balanced homodyne detection in the optics framework \cite{mistagauss}. Given a bipartite Gaussian state $\rho_{AB}$, any such  measurement on, say,  the $N_B$-mode subsystem $B=(B_1\ldots B_{N_B})$, is described by a POVM  of the form \cite{fiurasek07}
\begin{equation}\label{gpovm}
\Pi_B(\gr{\eta}) = \pi^{-N_B} \left[\prod_{j=1}^{N_B}  \hat{D}_{B_j}(\eta_j)\right] \Lambda^\Pi_B \left[\prod_{j=1}^{N_B}\hat{D}^\dagger_{B_j}(\eta_j)\right]\,,
 \end{equation}
 where
 \begin{equation}\label{weyl}
 \hat{D}_B(\eta_j) = \exp(\eta_j \hat{b}_j^\dagger - \eta_j^\ast \hat{b}_j)
  \end{equation} is the Weyl operator (\ref{weyl0}),
  $\hat{b}_j$ is the annihilation operator of the $j$-th mode of the subsystem $B$, $\pi^{-N_B}\int \Pi_B(\gr{\eta}) {\rm d}^{2N_B}\gr{\eta}  = \id$, and $\Lambda^\Pi_B$ is the density matrix of a (generally mixed) $N_B$-mode Gaussian state with CM $\gr{\Gamma}_B^\Pi$ which denotes the so-called seed of the measurement. The conditional state $\rho_{A|\gr{\eta}}$ of subsystem $A$ after the measurement $\Pi_B(\gr{\eta})$ has been performed on $B$ has a CM $\tilde{\gr\sigma}^{\Pi}_A$ independent of the outcome $\gr{\eta}$ and given by the Schur complement \cite{giedkefiurasekdistill}
\begin{equation}\label{eq:schur}
\tilde{\gr\sigma}^{\Pi}_A = \sig_A-\gr\varsigma_{AB} (\gr\sigma_B+ \gr\Gamma_B^\Pi)^{-1} \gr\varsigma_{AB}^{\sf T}\,,
 \end{equation}
 where the original bipartite CM $\sig_{AB}$ of the $N$-mode  state $\rho_{AB}$ has been written in block form as in Eq.~(\ref{eq:cms}).

\subsection{Gaussian information measures in terms of R\'enyi-$2$ entropy}
An extensive account of informational and entanglement properties of Gaussian states, using various well-established measures, can be found for instance in \cite{eisertplenio,ourreview,extremal,geof}. Here we follow a novel approach introduced in Ref.~\cite{renyi}, to which the reader is referred for further details and rigorous proofs.

R\'{e}nyi-$\alpha$ entropies \cite{Arenyi} constitute a powerful family of additive entropies, which provide a generalised spectrum of measures of (lack of) information in a quantum state $\rho$. They find widespread application in quantum information theory (see \cite{renyi} and references therein), while their role in holographic theories is attracting a certain interest from the gravity community as well \cite{holorenyi}. They are defined as
\begin{equation}\label{eq:ren}
{\cal S}_\alpha(\rho) = \frac1{1-\alpha} \ln \text{tr} (\rho^\alpha)\,,
\end{equation}
and reduce to the conventional von Neumann entropy in the limit $\alpha \rightarrow 1$.
The case $\alpha=2$ is especially simple,  ${\cal S}_2(\rho) = - \ln \text{tr} (\rho^2)\,.$

For arbitrary Gaussian states, the R\'enyi entropy of order $2$ satisfies the strong subadditivity inequality \cite{renyi}; this allows us to define relevant {\it bona fide} Gaussian measures of information and correlation quantities, encompassing entanglement and more general quantum and classical correlations, under a unified approach.

\paragraph{Mixedness.} For a Gaussian state $\rho$ with CM $\sig$, our preferred measure of mixedness (lack of purity, or, equivalently, lack of information, i.e., ignorance) will thus be  the R\'enyi-$2$ entropy,
\begin{equation}\label{eq:renyg}
{\cal S}_2(\sig) =  \frac12 \ln (\det \boldsymbol{\sigma})\,,
\end{equation}
which is $0$ on pure states ($\det \boldsymbol{\sigma}=1$) and grows unboundedly with increasing mixedness of the state.
This measure is directly related to the phase-space Shannon entropy of the Wigner distribution $W_{\rho}$ of the state $\rho$  (\ref{eq:wigner}), defined as $H(W_{\rho}(\boldsymbol{\xi}))=-\int W_{\rho}(\boldsymbol{\xi}) \ln\{W_{\rho}(\boldsymbol{\xi})\} {\rm d}^{2N}\boldsymbol{\xi}$ \cite{shan}. Indeed, one has $H(W_{\rho}(\boldsymbol{\xi}))={\cal S}_2(\rho)+N(1+\ln \pi)$ \cite{renyi}.

\paragraph{Total correlations.} For a bipartite Gaussian state $\rho_{AB}$ with CM as in \eq{eq:cms}, the  total correlations between subsystems $A$ and $B$ can be quantified by the R\'enyi-2 mutual information ${\cal I}_2$,
defined as \cite{renyi}
\begin{eqnarray}\label{eq:remutual}
{\cal I}_2(\sig_{A:B}) &=& {\cal S}_2(\sig_A) + {\cal S}_2(\sig_B) - {\cal S}_2(\sig_{AB}) \nonumber \\
&=&\frac12 \ln\left(\frac{\det\boldsymbol{\sigma}_A \det\boldsymbol{\sigma}_B}{\det\boldsymbol{\sigma}_{AB}}\right)\,,
\end{eqnarray}
which measures the phase space distinguishability between the Wigner function of $\rho_{AB}$ and the Wigner function associated to the product of the marginals $\rho_A \otimes \rho_B$, which is, by definition, a state in which the subsystems $A$ and $B$ are completely uncorrelated.

\paragraph{Entanglement.}
In the previous paragraph, we talked about total correlations, but that is not the end of the story. In general, we can discriminate between classical and quantum correlations. A bipartite pure state $|\psi_{AB}\rangle$ is quantum-correlated, i.e., is `entangled', if and only if it cannot be factorized as $|\psi_{AB}\rangle =|\phi_A\rangle\otimes |\chi_B\rangle$.   On the other hand, a mixed state $\rho_{AB}$ is entangled if and only if it cannot be written as $\rho_{AB}=\sum_i p_i \varrho_{A_i}\otimes \varrho_{B_i}$, that is a convex combinations of product states, where $\{p_i\}$ are probabilities and $\sum_i p_i=1$. Unentangled states are called `separable'. The reader can refer to Ref.~\cite{entanglementreview} for an extensive review on entanglement. In particular, one can quantify the amount of entanglement in a state by building specific measures. For Gaussian states, any measure of entanglement will be a function of the elements of the CM only \cite{ourreview}.

A measure of bipartite entanglement ${\cal E}_2$  for Gaussian states based on R\'enyi-$2$ entropy can be defined as follows \cite{renyi}. Given a  Gaussian state $\rho_{AB}$ with CM $\sig_{AB}$, we have
\begin{equation}\label{eq:GR2_ent}
{\cal E}_2 (\sig_{A:B}) = \inf_{\left\{\gam_{AB}\ :\ 0<\gam_{AB} \le \sig_{AB}, \, \det{\gam_{AB}}=1\right\}} \frac12 \ln \left(\det \gam_A\right)\,,
\end{equation}
where the minimisation is over pure $N$-mode Gaussian states with CM $\gam_{AB}$ smaller than $\sig_{AB}$.
For a pure Gaussian state $\rho_{AB} = \ket{\psi_{AB}}\bra{\psi_{AB}}$ with CM $\sig_{AB}^{\rm pure}$, the minimum is saturated by $\gam_{AB}=\sig_{AB}^{\rm pure}$, so that the measure of \eq{eq:GR2_ent} reduces to the pure-state R\'enyi-$2$ entropy of entanglement,
\begin{equation}\label{eq:GR2_ent_pure}
{\cal E}_2(\sig_{A:B}^{\rm pure})= {\cal S}_2(\sig_A) = \frac12 \ln (\det{\sig_A})\,,
\end{equation}
where $\sig_A$ is the reduced CM of subsystem $A$.
For a generally mixed state, Eq.~(\ref{eq:GR2_ent})  amounts to taking the Gaussian convex roof of the pure-state R\'{e}nyi-$2$ entropy of entanglement, according to the formalism of \cite{geof}. Closed formulae for ${\cal E}_2$ can be obtained for special classes of two-mode Gaussian states \cite{renyi}. The R\'enyi-$2$ entanglement is additive and monotonically nonincreasing under Gaussian local operations and classical communication.

\paragraph{Classical correlations.}
For pure states, entanglement is the only kind of quantum correlations. A pure separable state is essentially classical, and the subsystems display no correlation at all. On the other hand, for mixed states, one can identify a finer distinction between classical and quantum correlations, such that even most separable states display a definite quantum character \cite{zurek,vedral}.

Conceptually, one-way classical correlations are those extractable by local measurements; they can be defined in terms of how much the ignorance about the state of a subsystem, say $A$, is reduced when the most informative local measurement is performed on subsystem $B$ \cite{vedral}. The quantum correlations (known as `discord') are, complementarily, those destroyed by local measurement processes, and correspond to the change in total correlations between the two subsystems, following the action of a minimally disturbing local measurement on one subsystem only \cite{zurek}. For Gaussian states, R\'enyi-$2$ entropy can be adopted once more to measure ignorance and correlations \cite{renyi}.

To begin with, we can introduce a Gaussian R\'enyi-$2$ measure of one-way classical correlations \cite{vedral,giordaparis,adessodatta,renyi}. We define ${\cal J}_2(\sig_{A|B})$\footnote{\label{notesymm}Notice the directional notation ``$A|B$'' to indicate ``$A$ {\it given} $B$'', i.e., to specify that we are looking at the change in the informational content of $A$ following a minimally disturbing marginal  measurement on $B$. For entanglement and total correlations there is no direction as those quantities are symmetric, so the notation ``$A:B$'' is adopted instead.} as the maximum decrease in the R\'{e}nyi-$2$ entropy of subsystem $A$, given a Gaussian measurement has been performed on subsystem $B$, where the maximisation is over all Gaussian measurements [see Eqs.~\ref{gpovm},(\ref{eq:schur})].
We have then
\begin{eqnarray}\label{eq:J2}
{\cal J}_2(\sig_{A|B}) &=& \sup_{\gr\Gamma_B^\Pi} \frac12 \ln \left(\frac{\det \sig_A}{\det\tilde{\gr\sigma}^{\Pi}_A}\right)\,; \nonumber \\ & & \\
{\cal J}_2(\sig_{B|A}) &=& \sup_{\gr\Gamma_A^\Pi} \frac12 \ln \left(\frac{\det \sig_B}{\det\tilde{\gr\sigma}^{\Pi}_B}\right)\,, \nonumber
\end{eqnarray}
where the one-way classical correlations ${\cal J}_2(\sig_{B|A})$, with Gaussian measurements on $A$, have been defined accordingly by swapping the roles of the two subsystems, $A \leftrightarrow B$. Notice that, for the same state $\rho_{AB}$,  ${\cal J}_2(\sig_{A|B}) \neq {\cal J}_2(\sig_{B|A})$ in general: The classical correlations depend on which subsystem is measured\footref{notesymm}.

\paragraph{Quantum correlations.}
We can now define a Gaussian measure of quantumness of correlations based on R\'{e}nyi-$2$ entropy. Following the landmark study by Ollivier and Zurek \cite{zurek}, and the recent investigations of Gaussian quantum discord \cite{giordaparis,adessodatta,renyi}, we define the R\'enyi-$2$ discord as the difference between mutual information (\ref{eq:remutual}) and classical correlations (\ref{eq:J2}),
\begin{eqnarray}\label{eq:D2}
{\cal D}_2(\sig_{A|B}) &=& {\cal I}_2(\sig_{A:B})- {\cal J}_2(\sig_{A|B}) \nonumber\\
&=&\inf_{\gr\Gamma_B^\Pi} \frac12 \ln \left(\frac{\det \sig_B \det \tilde{\gr\sigma}^{\Pi}_A}{\det \sig_{AB}}\right)\,;
\nonumber \\ & & \\
{\cal D}_2(\sig_{B|A}) &=& {\cal I}_2(\sig_{A:B})- {\cal J}_2(\sig_{B|A}) \nonumber \\
&=&\inf_{\gr\Gamma_A^\Pi} \frac12 \ln \left(\frac{\det \sig_A \det \tilde{\gr\sigma}^{\Pi}_B}{\det \sig_{AB}}\right)\,.
\nonumber
\end{eqnarray}
The discord is clearly a nonsymmetric quantity as well\footref{notesymm}. It captures general quantum correlations even in the absence of entanglement \cite{zurek,modireview}.

Let us remark that we have defined classical and quantum correlations by restricting the optimisation over Gaussian measurements only. This means that, potentially allowing for more general non-Gaussian measurements, one could obtain higher classical correlations and lower quantum ones. However, some numerical and partial analytical evidence support the conclusion that, for two-mode Gaussian states, Gaussian measurements are optimal for the calculation of general one-way classical and quantum discord \cite{adessodatta,allegramente}. Certainly, restricting to the practically relevant Gaussian measurements makes the problem dramatically more tractable, as one can obtain closed analytical expressions for Eqs.~(\ref{eq:J2}) and (\ref{eq:D2}) for the case of $A$ and $B$ being single modes, that is, $\rho_{AB}$ being a general two-mode Gaussian state \cite{adessodatta,renyi}. We make explicit use of these formulae to derive the results of Sec.~\ref{secalice}. Further, restricting to Gaussian measurements also corresponds pretty much to the reality of what is implementable in laboratory with present day technology \cite{expdiscordgauss}.

Finally, let us observe that \begin{eqnarray}\label{pureeq}
\frac{{\cal I}_2(\sig^{\rm pure}_{A:B})}2&=&{\cal J}_2(\sig^{\rm pure}_{A|B})={\cal J}_2(\sig^{\rm pure}_{B|A})={\cal D}_2(\sig^{\rm pure}_{A|B})={\cal D}_2(\sig^{\rm pure}_{B|A})\nonumber \\ &=&{\cal E}_2(\sig^{\rm pure}_{A:B})={\cal S}_2(\sig_A)={\cal S}_2(\sig_B)\,,
\end{eqnarray}
for {\it pure} bipartite Gaussian states $\rho_{AB}$ of an arbitrary number of modes. That is, general quantum correlations reduce to entanglement, and an equal amount of classical correlations is contained as well in pure states.

\section{Unruh effect and correlations of scalar field modes in noninertial frames}\label{secalice}

\subsection{Rudiments of the Unruh effect}

There are excellent references in the literature about the Unruh effect \cite{unruh}, see e.g.~\cite{unruhreview} for a recent review; the physics of it will be most likely covered in detail elsewhere in this Special Issue. We shall briefly recall the phenomenon for the purpose of setting up our notation.

It is well known that different quantisation procedures for observers in different states of motion, i.e., inertial and noninertial observers, of a quantum field in a flat spacetime may introduce not only non-trivial effects on particle generation, but also on the behaviour of the correlations between field modes. The setting we wish to investigate is the following. We consider a $(1+1)$-dimensional Minkowski spacetime with coordinates $(t,z)$, which we can adopt as proper coordinates for an inertial observer Alice moving in the Minkowski plane. In such a context, the proper coordinates of an observer Rob moving  with uniform proper acceleration $a$  are the Rindler coordinates $(\tau,\zeta)$. Two different sets of Rindler coordinates are needed for covering region $I, II$ of the Minkowski spacetime (see Fig.~\ref{figallo}),  and are given by
\begin{eqnarray}\label{Rindler_coords}
I &:& \quad a t =  {\rm e}^{a\zeta}\sinh(a\tau), \quad a z =
{\rm e}^{a\zeta}\cosh(a\tau),\\
II &:& \quad a t =
-{\rm e}^{a\zeta}\sinh(a\tau), \quad a z =
-{\rm e}^{a\zeta}\cosh(a\tau).
\end{eqnarray}
These sets of coordinates define two Rindler regions (respectively
$I$ and $II$) that are causally disconnected from each other.

Now, let us consider a free quantum scalar field: Its quantisation in the Minkowski coordinates is not equivalent to the one in the Rindler ones, since the solutions of the Klein-Gordon equation in the two coordinate systems are different. In particular, a Minkowski vacuum state of a field mode described by an inertial observer Alice is expressed in Rindler coordinates as a two-mode squeezed state:
\begin{eqnarray}
|0_{\rm M}\rangle=\hat{U}_{I,II}(r)|n\rangle_I|n\rangle_{II}=\frac{1}{\cosh r}\sum_{n=0}^{\infty}\tanh^n r |n\rangle_I|n\rangle_{II},
\end{eqnarray}
where $\hat{U}_{I,II}(r)$ is exactly the two-mode squeezing operator of \eq{tmsU},  that encodes the particle pair production between the two Rindler wedges. Here
the dimensionless `acceleration parameter' $r$ is proportional to the Unruh temperature $T$:
\begin{equation}\label{Tunruh}
  \cosh^{-2} r = 1-{\rm e}^{-\frac{\hbar |\omega|}{k_B T}}\,,\quad T=\frac{\hbar a}{2\pi k_B }\,,
\end{equation}
with $k_B$ being the Boltzmann constant, and $\omega$ being the frequency of the mode.

Adopting the Heisenberg picture, we have that the Rindler field mode operators $\hat{b}_{I,II}$ are connected to the Minkowski ones $\hat{b}_{\rm M}$ via a Bogoliubov transformation \cite{birelli},
\begin{equation}\label{bogo}
\hat{b}_{\rm M}=\cosh r\ \hat{b}_{I} - \sinh r\ \hat{b}^{\dagger}_{II}\,.
\end{equation}

A noninertial observer Rob with uniform acceleration $a$ is confined to Rindler region $I$ and has no access to the opposite region.  Thus the equilibrium state from Rob's viewpoint, in the Schr\"odinger picture, is obtained by tracing over the modes in the causally disconnected region $II$,
 \begin{eqnarray}\label{stator}
 \rho_I &=& {\rm tr}_{II} \left\{\hat{U}_{I,II}(r) \big[(\ket{0_{\rm M}}\!\bra{0_{\rm M}})_I \otimes (\ket{0}\!\bra{0})_{II}\big] \hat{U}_{I,II}^\dagger(r)\right\} \nonumber \\
 &=&\frac{1}{\cosh^2 r}\sum_{n=0}^{\infty}\tanh^{2n} r|n\rangle\langle n|_I\,.
 \end{eqnarray}
One can then see that the Minkowski vacuum is described, by a uniformly accelerated observer Rob, as a particle-populated thermal state with temperature $T$ given by \eq{Tunruh}

 This phenomenon, called Unruh effect \cite{unruh}, has a well known formal analogue in quantum optics \cite{barnett}:  An input signal beam $I$ in the state $\ket{0_{\rm M}}$ interacts with an idler vacuum mode $II$ (ancilla) via a two-mode squeezing transformation $\hat{U}_{I,II}(r)$ (realised by parametric down-conversion) with squeezing $r$; tracing over the output idler mode, the output signal is left precisely in the mixed thermal state $\rho_I$ of \eq{stator}. Overall the non-unitary transformation from input to output, or from inertial to noninertial frame, corresponds to the action of a {\it bosonic amplification channel} \cite{bradler,mariona}.

 One can question how a different state (other than the vacuum) of a scalar field mode, described as $\ket{\psi}$ in Minkowski coordinates, is perceived by a noninertial observer Rob confined to Rindler region $I$. In seminal RQI investigations \cite{alicefalls, alicefermion, unruhsharing}, it was implicitly assumed that a Minkowski mode with a sharp frequency transforms into a single frequency Rindler mode too. This assumption has been proven incorrect \cite{bruschi}: Minkowski modes prepared in states other than the vacuum, e.g.~single-particle states, are effectively described as oscillatory, non-peaked broadband wavepackets from a Rindler perspective. However, a valid `single-mode approximation' can be still employed if one considers in general a class of Unruh modes \cite{unruh,birelli} of the massless scalar field, rather than Minkowski modes.  Such modes are purely positive-frequency combinations of standard plane waves in Minkowski coordinates, but enjoy a special property: they are mapped into single frequency modes in Rindler coordinates. The interested reader can refer e.g.~to \cite{bruschi,eduunruh} for further details. Unruh modes form a complete basis of solutions of the field equations that can span any physical state, and are very apt to make calculations, therefore qualifying as suitable candidates for our exploratory investigation. Although they have been shown to suffer some pathologies (delocalisation and oscillatory behaviour near the acceleration horizon) that might hinder their physical realisation \cite{bruschi,nathan}, one can in principle design plausible models of non-point-like detectors which couple effectively to a single Unruh mode \cite{squash}.

Having clarified these important issues of physical nature, let us recall the mathematical results. Let $\ket{\psi}_{\rm U}$ denote the state of a Unruh mode of the field from an inertial perspective, characterised by the creation operator
\begin{eqnarray}\label{brusci}
\hat{b}^\dagger_{\rm U} \ket{0_{\rm M}} &=& \frac{1}{\cosh^2 r}\sum_{n=0}^{\infty}\tanh^n r \sqrt{n+1} \ket{\Phi_n}\,, \nonumber \\
\mbox{with }\ket{\Phi_n} &=& q_L \ket{n}_I \ket{n+1}_{II} + q_R \ket{n+1}_I \ket{n}_{II}\,,
\end{eqnarray}
where $|q_R|^2+|q_L|^2=1$ and $\ket{\Phi_n}$ is the state of a single Rindler mode of frequency $\omega$, see \cite{bruschi} for details. If we fix $q_R=1$, $q_L=0$, then the formal analogy with the bosonic amplification channel still holds for any inertial state $\ket{\psi}_{\rm U}$ of the Unruh mode, and a formula akin to Eq.~(\ref{stator}) can be still used to determine the state $\rho_I$ of the field as described in Rindler coordinates by the noninertial observer Rob \cite{birelli,bruschi,mariona}. One has, namely
 \begin{eqnarray}\label{stator2}
 \rho_I &=& {\rm tr}_{II} \left\{\hat{U}_{I,II}(r) \big[(\ket{\psi}\!\bra{\psi})_I \otimes \ (\ket{0}\!\bra{0})_{II}\big] \hat{U}_{I,II}^\dagger(r)\right\}\,,
 \end{eqnarray}
where the Rindler modes $I$ and $II$ have a definite frequency $\omega$. If $\ket{\psi}$ is the Unruh vacuum, which coincides with the Minkowski vacuum $\ket{0_{\rm M}}$, then \eq{stator2} reduces to \eq{stator}. In general, $\rho_I$ will be some other mixed state from a noninertial perspective. We remark that one might choose different values of $q_{R,L}$ in \eq{brusci}, which could result in interesting phenomena such as enhancement rather than degradation of quantum correlations from noninertial perspectives \cite{amplif}; we leave those settings for further analysis, focusing here on the case $q_R=1$.

  The crucial observation to make is that the two-mode squeezing transformation $\hat{U}_{I,II}(r)$ is a Gaussian operation, and the amplification channel is a Gaussian channel, i.e., they preserve the Gaussianity of the input states. Therefore, if the inertial state $\ket{\psi}$ is chosen to be Gaussian in \eq{stator2}, the transformed states in Rindler coordinates remain Gaussian as well, and the methods from the previous Section can be readily employed to characterise how informational properties are perceived in different reference frames \cite{unruhsharing}. This holds for general Bogoliubov transformations \cite{niconew}.

\subsection{The setting}

In this paper we focus on a massless scalar field whose state, as seen from an inertial Minkowski frame, involves all modes in the vacuum, but for two Unruh modes $A$ and $R$ which are initialised in a pure, entangled (Gaussian) two-mode squeezed state with squeezing $s$ \cite{unruhsharing,ahnkim}, characterised by a CM of the form
\begin{eqnarray}\label{inAR}
\sig^{\rm (M)}_{AR}(s)=\gr S_{A,R}(s)\id_{AR}\gr S_{A,R}\T(s) = \left(\!\!\begin{array}{cccc}
\cosh (2s)&0&\sinh (2s)&0\\
0&\cosh (2s)&0&-\sinh (2s)\\
\sinh (2s)&0&\cosh (2s)&0\\
0&-\sinh (2s)&0&\cosh (2s)
\end{array}\!\!\right),\nonumber \\
\end{eqnarray}
where $\id_{AR}$ is the CM of the two-mode vacuum $\ket{0_{\rm M}}_{k_A}\!\otimes\!\ket{0_{\rm M}}_{k_R}$, and $\gr S_{i,j}$  defined in \eq{tmsS} is the phase space (symplectic) representation of the two-mode squeezing operation of \eq{tmsU}. Notice that, compared with the previous subsection, we are considering here two populated Unruh modes (rather than a single one) which are already correlated from an inertial perspective.

Let us then first analyse the correlations between the two modes $A$ and $R$ in Minkowski coordinates. Let Alice be the observer associated to the description of the mode $A$, and let Rob be the observer who describes mode $R$. In our analysis, we are avoiding the issues associated to the practical detection of those modes: The reader can consider the terms `observers' and `coordinates' as synonyms for all practical purposes.
From a fully inertial perspective (i.e., if both observers are inertial), the correlations in the state $\sig^{\rm (M)}_{AR}(s)$ are given by Eq.~(\ref{pureeq}), that is, entanglement, quantum and classical correlations are all equal to
\be\label{incorr}
{\cal C}_2\big(\sig^{\rm (M)}_{A:R}(s)\big) \equiv \ln[\cosh(2s)]
\ee
(where we have introduced the common symbol ${\cal C}_2$ for `correlations'),
while total correlations are clearly ${\cal I}_2 =2 {\cal C}_2$. All the correlations increase unboundedly with increasing initial squeezing $s$. For $s \rightarrow \infty$ the two modes of the field asymptotically tend to realise, from an inertial perspective, the Einstein-Podolski-Rosen perfectly correlated state.

\begin{figure}[t]
\begin{center}
\subfigure[]{\includegraphics[width=7cm]{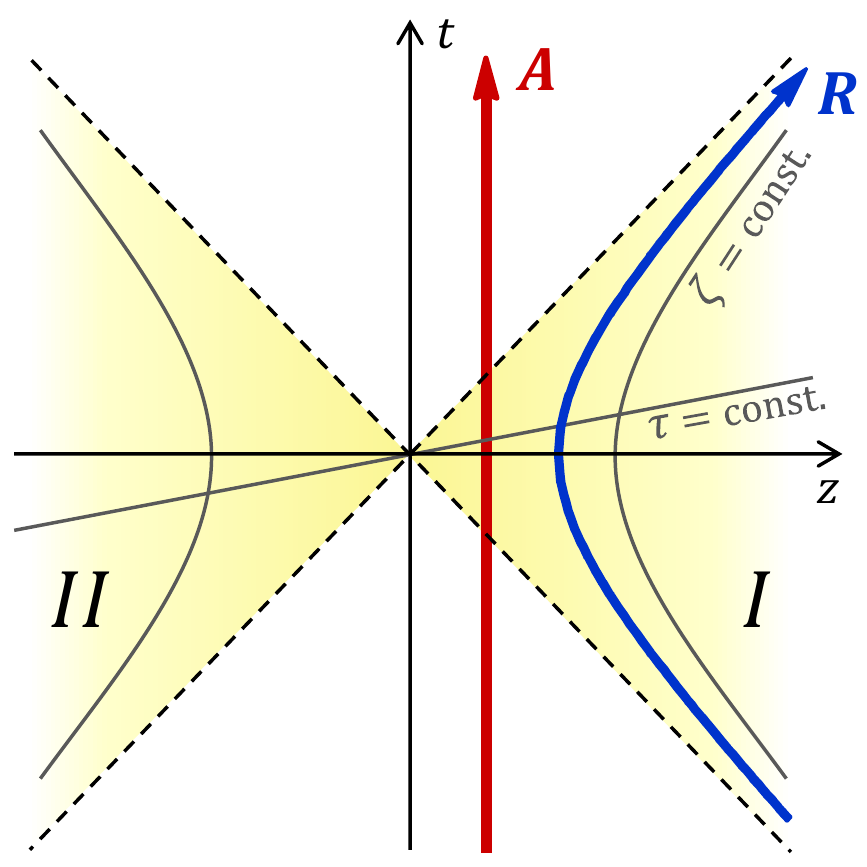}}
\hspace*{1cm}
\subfigure[] {\includegraphics[width=7cm]{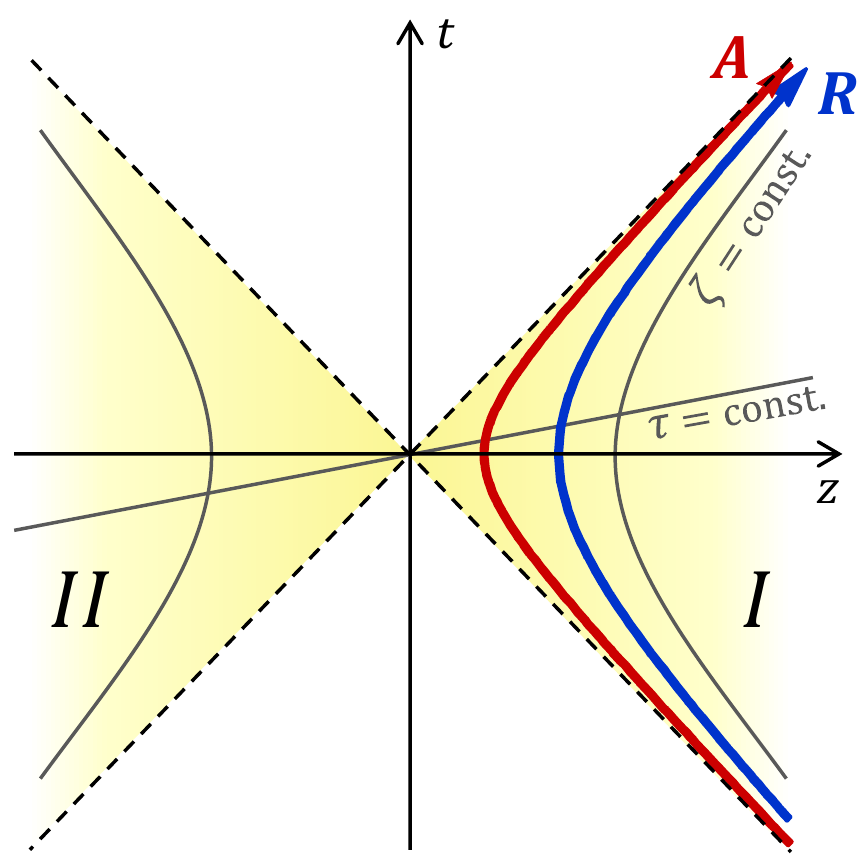}}
\caption{(Color online)
Sketch of the world lines for two observers Alice and Rob in two different settings: (a) Alice inertial, Rob uniformly accelerated; (b) both observers in uniform acceleration.
The set $(z,t)$ denotes Minkowski coordinates, while the set
$(\zeta, \tau)$ denotes Rindler coordinates. The causally
disconnected Rindler regions $I$ and $II$ are evidenced. \label{figallo}}
\end{center}
\end{figure}

Now, we present the core of our work. We shall consider two settings. In the first one, say (a),  the mode $A$ is still expressed in Minkowski coordinates, i.e., Alice is an inertial observer, while $R$ is now described by Rindler coordinates, i.e., Rob undergoes uniform acceleration characterised by an acceleration parameter $r$. In the second picture, say (b), Alice and Rob are both subjected to uniform acceleration characterised by acceleration parameters $w$ and $r$, respectively. The world lines for the two settings are depicted schematically in Fig.~\ref{figallo}.
Entanglement redistribution phenomena under these prescriptions have been studied for instance in \cite{unruhsharing,buchi} for scalar fields.

For setting (a), the complete description of the problem involves three modes, mode $A$ described by the inertial Alice, mode $R$  by the noninertial Rob in Rindler region $I$, and mode $\bar{R}$  by a noninertial observer anti-Rob confined to Rindler region $II$  (the prefix  `anti' is just used for labelling observers in region $II$). This is because the mode $R$ is mapped to two sets of Rindler coordinates, respectively for region $I$ and $II$. Consistently, setting (b) involves additionally a fourth mode, $\bar{A}$, as we now  have a noninertial observer anti-Alice confined to Rindler region $II$ as well.

Let us now analyse the description of the two settings in the Gaussian phase space formalism. Reminding we can work at the CM level, we have already seen from \eq{stator2} that the change from Minkowski to Rindler coordinates corresponds to a two-mode squeezing operation for each single Unruh mode, i.e.,
$|\psi\rangle_{\rm U}=\hat{U}_{I,II}(r)(|\psi\rangle_I\otimes|0\rangle_{II}),$
where $\hat{U}_{I,II}(r)$ is associated to the symplectic transformation $\gr S_{I,II}(r)$.

In the first setting, therefore, since the observer Rob is accelerating uniformly, the original two-mode entangled states described by Alice and Rob in Minkowski coordinates becomes `distributed' among three observers, i.e., we need three systems of coordinates to describe it, specifically associated to Alice, Rob for the region $I$ and anti-Rob for region $II$. The Gaussian state of the complete system has CM given by \cite{unruhsharing}
\begin{eqnarray}\label{in34}
\sig^{\rm (a)}_{AR \bar R}(s,r) &=& \big[\id_A \oplus \gr S_{R,\bar R}(r)\big] \big[\sig^{\rm (M)}_{AR}(s) \oplus \id_{\bar R}\big] \big[\id_A \oplus \gr S_{R,\bar R}(r)\big]\T\,,
\end{eqnarray}
where $S_{R,\bar R}(r)$ is the squeezing operation correspondent to the change of coordinates due to the noninertiality of Rob, and we have used the fact that the CM of a vacuum state is the identity matrix.

Similarly, a change of coordinates for mode $A$ as well implies a further two-mode squeezing operation $\gr S_{A, \bar{A}}(w)$ where Alice is now a noninertial observer confined in region $I$ and anti-Alice in region $II$. Therefore, the CM for the complete state in the second setting reads
\begin{eqnarray}
\sig^{\rm (b)}_{AR \bar A \bar R}(s,w,r) &=& \big[\gr S_{A, \bar{A}}(w) \oplus \gr S_{R,\bar R}(r)\big] \big[\sig^{\rm (M)}_{AR}(s) \oplus \id_{\bar A\bar R}\big] \big[\gr S_{A, \bar{A}}(w) \oplus \gr S_{R,\bar R}(r)\big]\T\,\label{in4}.
\end{eqnarray}
Notice that if we set $w=0$ then mode $\bar{A}$ decouples and setting (b) reduces to setting (a).

\subsection{Correlations in noninertial frames}

\begin{figure}[t]
\begin{center}
{\includegraphics[width=11cm]{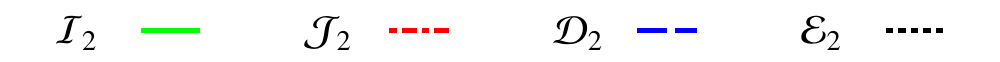}}\\
\vspace*{.2cm}
\subfigure[]{\includegraphics[width=7.7cm]{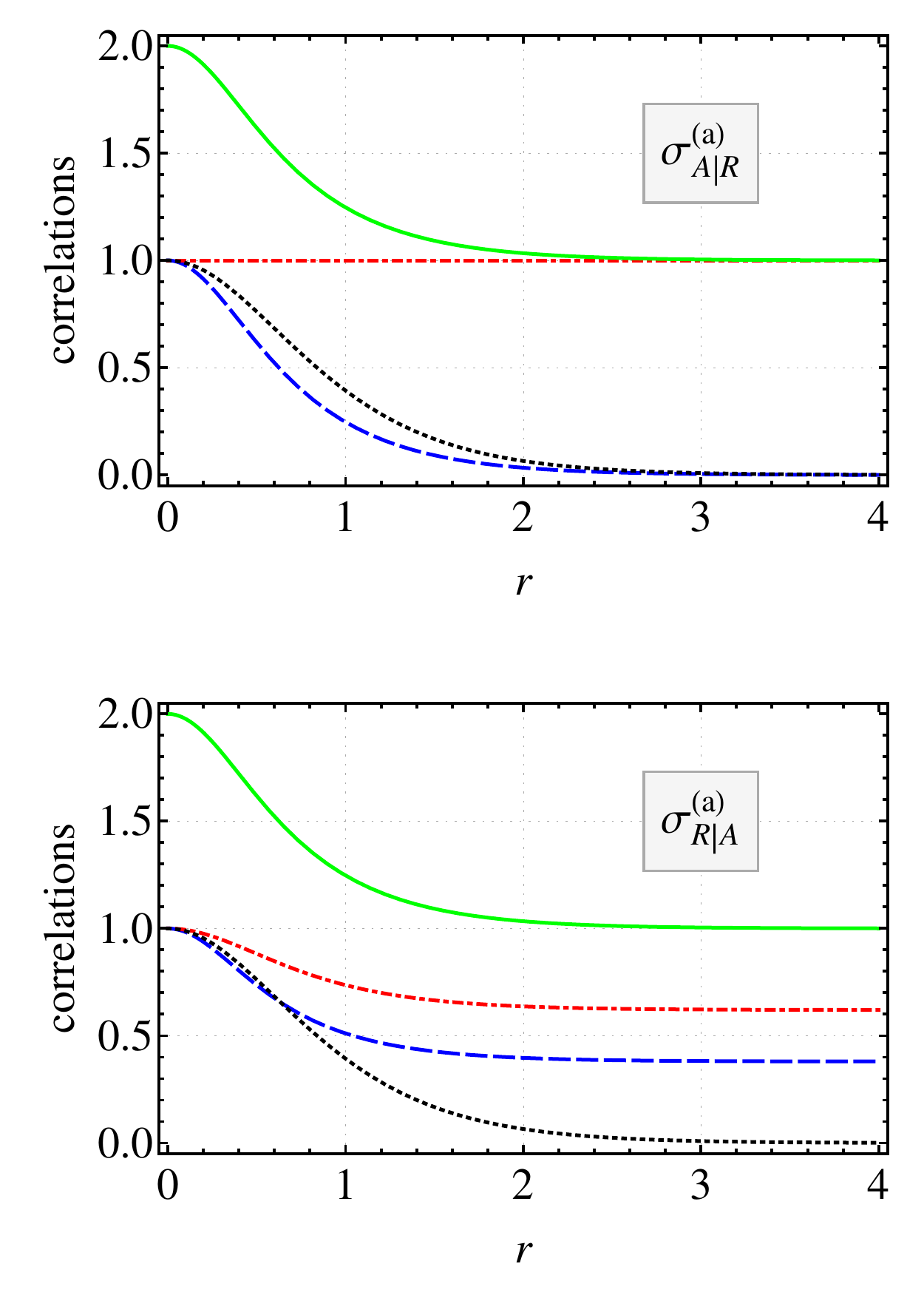}}
\subfigure[] {\includegraphics[width=7.7cm]{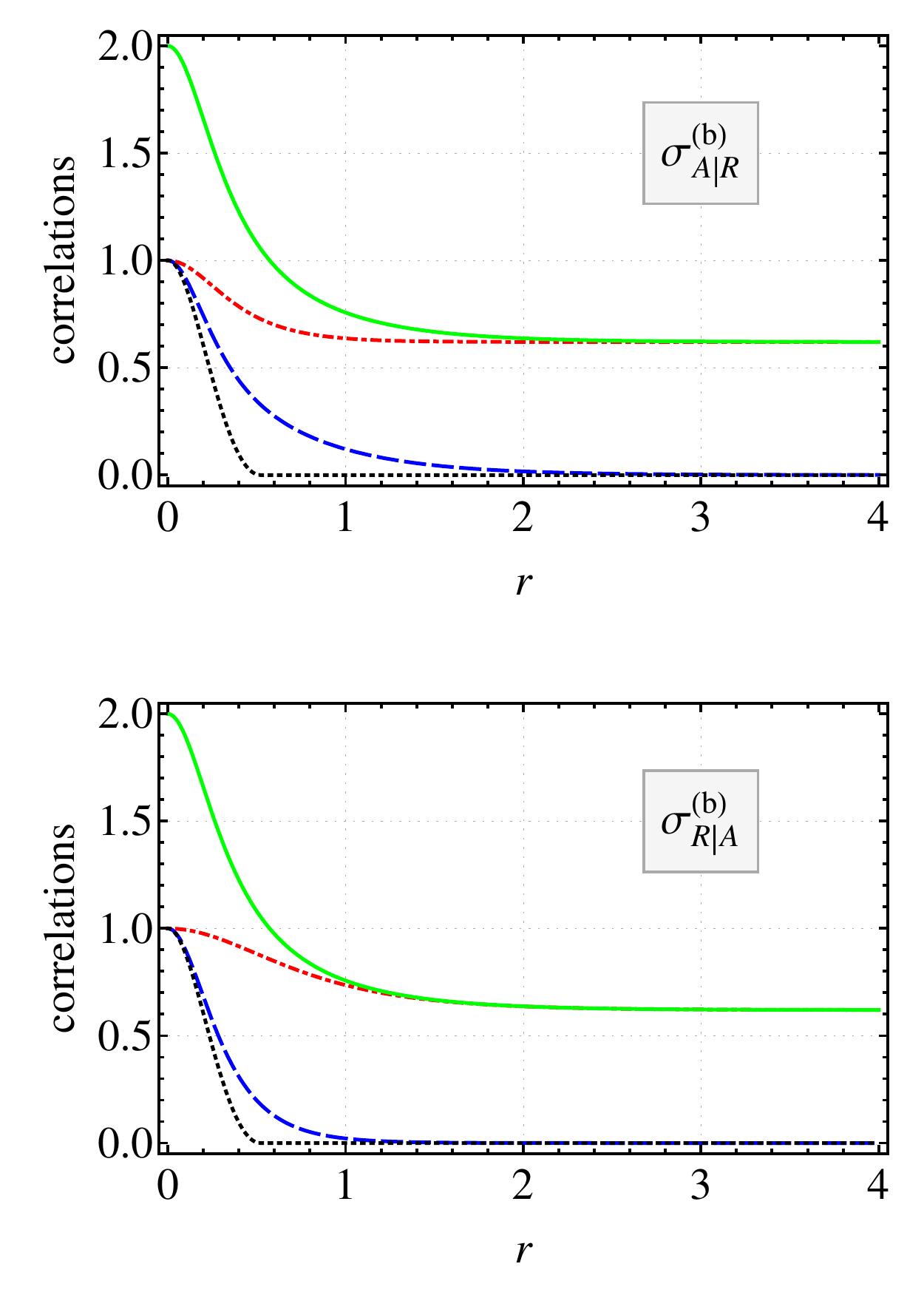}}
\caption{(Color online) Correlations between field modes $A$ and $R$ as described by Alice and Rob in the following settings. Column (a): Alice is inertial, and Rob undergoes uniform acceleration with acceleration parameter $r$. Column (b) Alice and Rob both undergo uniform acceleration, with parameters $w=2r$ and $r$ respectively. From a fully inertial perspective, the state of modes $A$ and $R$ is a two-mode squeezed state with squeezing parameter $s$ (we choose $s=\text{arccosh}({\rm e})/2$ in the plots, so that the inertial quantum and classical correlations amount to $1$). In the top row, classical and quantum correlations (quantified  respectively by the one-way measure ${\cal J}_2$ and the discord ${\cal D}_2$) are revealed through marginal measurements on $R$. In the bottom row, they are revealed through marginal measurements on $A$. Entanglement ${\cal E}_2$ and total correlations ${\cal I}_2$ are instead symmetric quantities. Notice how the classical correlations in the bottom row are the same for both settings. Notice also how entanglement vanishes at a finite $r$ for setting (b), while discord can only vanish in the infinite acceleration limit. Discord revealed through measurements on $A$ does not vanish in this limit when Alice stays inertial.\label{figafigra}}
\end{center}
\end{figure}

\paragraph{Correlations between physical observers.}
First of all, we are interested in the correlations between the two field modes $A$ and $R$ as described by the  observers Alice and Rob in the two settings described above. To obtain $\sig^{\rm (a,b)}_{AR}$, we need to trace Eqs.~(\ref{in34}),(\ref{in4}) over the unaccessible degrees of freedom associated to modes in Rindler region $II$. The latter modes appear to have acquired correlations with $A$ and $R$ from a noninertial perspective, as a consequence of the Unruh effect. Therefore, the physical state of modes $A$ and $R$ should be detected by Alice and Rob as more mixed and less correlated, intuitively, with increasing acceleration of one or both observers.
The correlations in the reduced states $\sig^{\rm (a,b)}_{AR}$ will be compared with the ones available from a fully inertial perspective, ${\cal C}_2\big(\sig^{\rm (M)}_{A:R}\big)$, given by \eq{incorr}. A comparative summary of our results is illustrated in Fig.~\ref{figafigra}.

We find {\it prima facie} that the total correlations [\eq{eq:remutual}] are degraded as functions of $r$ and $w$ as expected.
For the general setting (b), we get
\begin{eqnarray}\label{MIdeg}
{\cal I}_2(\sig^{\rm (b)}_{A:R})=\ln \left[\frac{\left(\cosh ^2(r) \cosh (2 s)+\sinh ^2(r)\right) \left(\cosh (2 s) \cosh ^2(w)+\sinh ^2(w)\right)}{\cosh (2 r)
   \cosh ^2(s) \cosh (2 w)-\sinh ^2(s)}\right], \nonumber \\
\end{eqnarray}
and the case of setting (a) can be retrieved by choosing $w=0$. The total correlations are never completely destroyed under the Unruh effect in the considered settings.
If Alice stays inertial [setting (a)], we get \be\lim_{r \rightarrow \infty} \frac{{\cal I}_2(\sig^{\rm (a)}_{A:R})}{{\cal I}_2\big(\sig^{\rm (M)}_{A:R}\big)}=\frac12\,;\ee while if Alice is in high uniform acceleration as well [setting (b)], we get \be\frac14\leq \lim_{r,w \rightarrow \infty} \frac{{\cal I}_2(\sig^{\rm (b)}_{A:R})}{{\cal I}_2\big(\sig^{\rm (M)}_{A:R}\big)}\leq \frac12\,.\ee

It is interesting to evaluate how the total correlations decompose into classical and genuinely quantum components.
We find that the classical correlations revealed in a marginal measurement are unaffected by the state of motion of the observer who is performing the measurement, but carry a signature of the state of motion of the other observer. Specifically, in setting (a), if the noninertial observer Rob implements a Gaussian measurement on mode $R$ and we compute the ensuing classical correlations [\eq{eq:J2}] between the modes $R$ and $A$, we find
\begin{equation}
{\cal J}_2(\sig^{\rm (a)}_{A|R}) = {\cal C}_2\big(\sig^{\rm (M)}_{A:R}\big)=\ln[\cosh(2s)]\,,
\end{equation}
independently of Rob's acceleration parameter $r$. In other words, if Alice stays inertial, she does not experience any lack of information on the state of mode $A$ depending on whether Rob detects $R$ in an inertial or a noninertial frame. We can then conclude, in this specific sense, that such one-way classical correlations are unaffected by the Unruh effect.
If we swap the roles over and let Alice be the one who implements the marginal measurement, however, we find instead that the classical correlations do depend on Rob's acceleration parameter $r$, yet they do not depend on Alice's acceleration parameter $w$: They are then the same in both settings (a) and (b) and given by
\begin{equation}
{\cal J}_2(\sig^{\rm (a,b)}_{R|A}) = \ln [\text{sech}(2 r) (\cosh ^2(r) \cosh (2 s)+\sinh ^2(r))]\,.
\end{equation}
For high acceleration of Rob ($r \gg 0$) and large inertial correlations ($s \gg 0$), we have
\begin{equation}\label{abit}
\lim_{r,s \rightarrow \infty} {\cal J}_2(\sig^{\rm (a,b)}_{R|A}) ={\cal C}_2\big(\sig^{\rm (M)}_{A:R}\big)-\ln 2\,,
 \end{equation} meaning that Rob experiences a lack of no more than one bit (which in our units amounts to $\ln 2$) of classical correlations even if Alice stays inertial, as predicted in \cite{unruhsharing}.

By combining the analysis of mutual information and classical correlations we can draw conclusions about the Unruh effect on general quantum correlations as quantified by the discord [\eq{eq:D2}]. We find that in setting (a), the discord revealed through measurements on $A$ converges to a finite value in the limit $r \rightarrow \infty$, given by
\begin{equation}\label{dabit}
\lim_{r \rightarrow \infty} {\cal D}_2(\sig^{\rm (a)}_{R|A}) = \ln \left[\frac{\cosh (2 s)}{\cosh^2 s}\right] {}_{\overrightarrow{{\ }_{s \rightarrow \infty}{\ }}} \ln2\,.
\end{equation}
This shows on one hand that not all genuinely quantum features are necessarily destroyed by the Unruh effect, and on the other hand that the loss of a bit of classical correlations, as shown in \eq{abit}, is somehow compensated, when Alice stays inertial, by the endurance of a bit of quantum discord, both revealed through marginal measurements on $A$. The permanence of a nonzero amount of discord in a similar context was found in \cite{dattaunruh} for scalar fields, albeit starting from a different (non-Gaussian) state for the modes $A$ and $R$ in a fully inertial perspective (in that case, the instance of measurements on $R$ could not be worked out).
Here we find that, for any other setting, namely setting (a) with measurements on $R$, and setting (b) for both directions, the discord goes asymptotically to zero when the involved acceleration parameters diverge.

Finally, let us just recall that entanglement between $A$ and $R$ also goes to zero asymptotically for $r \rightarrow \infty$ in setting (a) \cite{alicefalls,unruhsharing}, while it experiences so-called sudden death in setting (b) \cite{unruhsharing,buchi}, i.e., it can vanish for a range of finite values of the accelerations of Alice and Rob. Explicitly,
\begin{eqnarray}\label{ent2s}
{\cal E}_2(\sig_{A:R}^{\rm (a)})&=& \ln \left[\frac{(\cosh (2 r)+3) \cosh (2 s)+2 \sinh ^2(r)}{2 \sinh ^2(r) \cosh (2
   s)+\cosh (2 r)+3}\right]\,, \nonumber
\\ && \\
{\cal E}_2(\sig_{A:R}^{\rm (b)})&=&\left\{\!\!
                                    \begin{array}{l}
                                      0, \quad \hbox{if $\tanh s \leq \sinh w \sinh r$;}   \\
                                      \ln \left[\frac{-4 \sinh  w \sinh r \sinh (2 s)+2 \cosh (2 w) \cosh (2 r) \cosh
   ^2(s)+3 \cosh (2 s)-1}{2 \left(2 \sinh  w \sinh r \sinh (2 s)+\cosh ^2(s)
   (\cosh (2 w)+\cosh (2 r))-2 \sinh ^2(s)\right)}\right],  \hbox{ otherwise.}\!\!\!\!
                                    \end{array}
                                  \right. \nonumber
\end{eqnarray}

\paragraph{Correlations involving unaccessible modes.}
We wish to stress that the measures reported in Sec.~\ref{secgauss} allow us to calculate explicitly all forms of correlations and entanglement between all bipartitions in the complete states $\sig^{\rm (a)}_{AR \bar R}(s,r)$ and $\sig^{\rm (b)}_{AR \bar A \bar R}(s,w,r)$, involving also those modes $\bar A$ and $\bar R$ confined to the causally disconnected Rindler region $II$ and detected, in principle, by  observers anti-Alice and anti-Rob. For entanglement, this was done in \cite{unruhsharing} using different measures. Here, we do not have the space (and time) to adapt and extend such a study to encompass discord and classical correlations as well, although we believe this may constitute an interesting topic to expand upon elsewhere. We do, however, care to remark that the inertial entanglement between $A$ and $R$ (and the quantum correlations thereof) lost to the Unruh effect, can be partially interpreted and recovered in terms of {\it genuine} multipartite entanglement (respectively, discord) distributed among the accessible modes $A$, $R$, and the unaccessible ones $\bar{R}$ and $\bar{A}$ from noninertial perspectives. While this aspect was explored in \cite{unruhsharing,buchi}, the measures adopted in \cite{renyi} and in this paper are especially suited for such a task, as the R\'enyi-$2$ entanglement ${\cal E}_2$ satisfies a general `monogamy' inequality \cite{contangle} on entanglement sharing for multimode Gaussian states, and the R\'enyi-$2$ discord ${\cal D}_2$ enjoys the same property in the special case of pure three-mode Gaussian states \cite{renyi}, such as the states with CM $\sig_{AR\bar{R}}^{\rm (a)}$ [Eq.~(\ref{in34})].

\begin{figure}[t]
\begin{center}
\subfigure[]{\includegraphics[width=7.5cm]{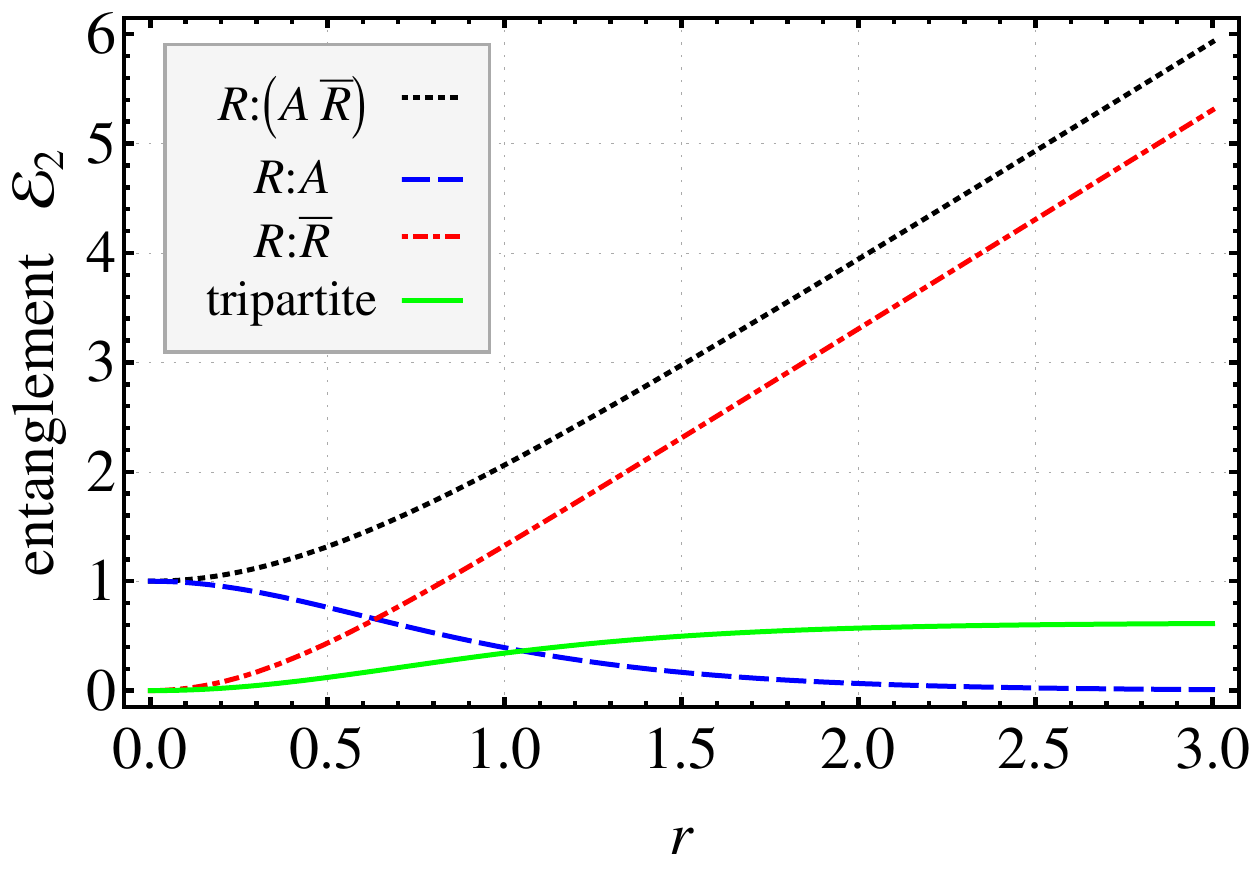}}\hspace*{0.4cm}
\subfigure[] {\includegraphics[width=7.5cm]{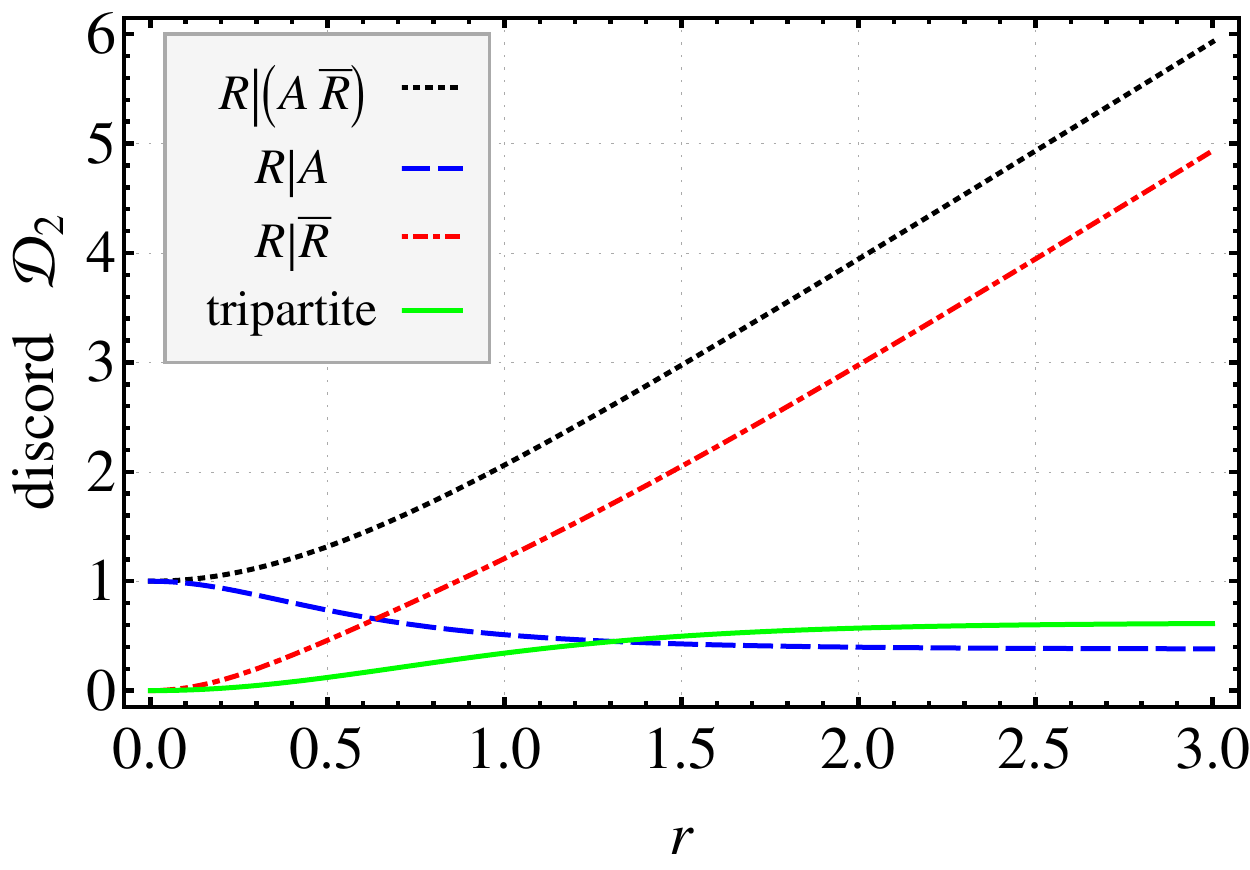}}
\subfigure[] {\includegraphics[width=9cm]{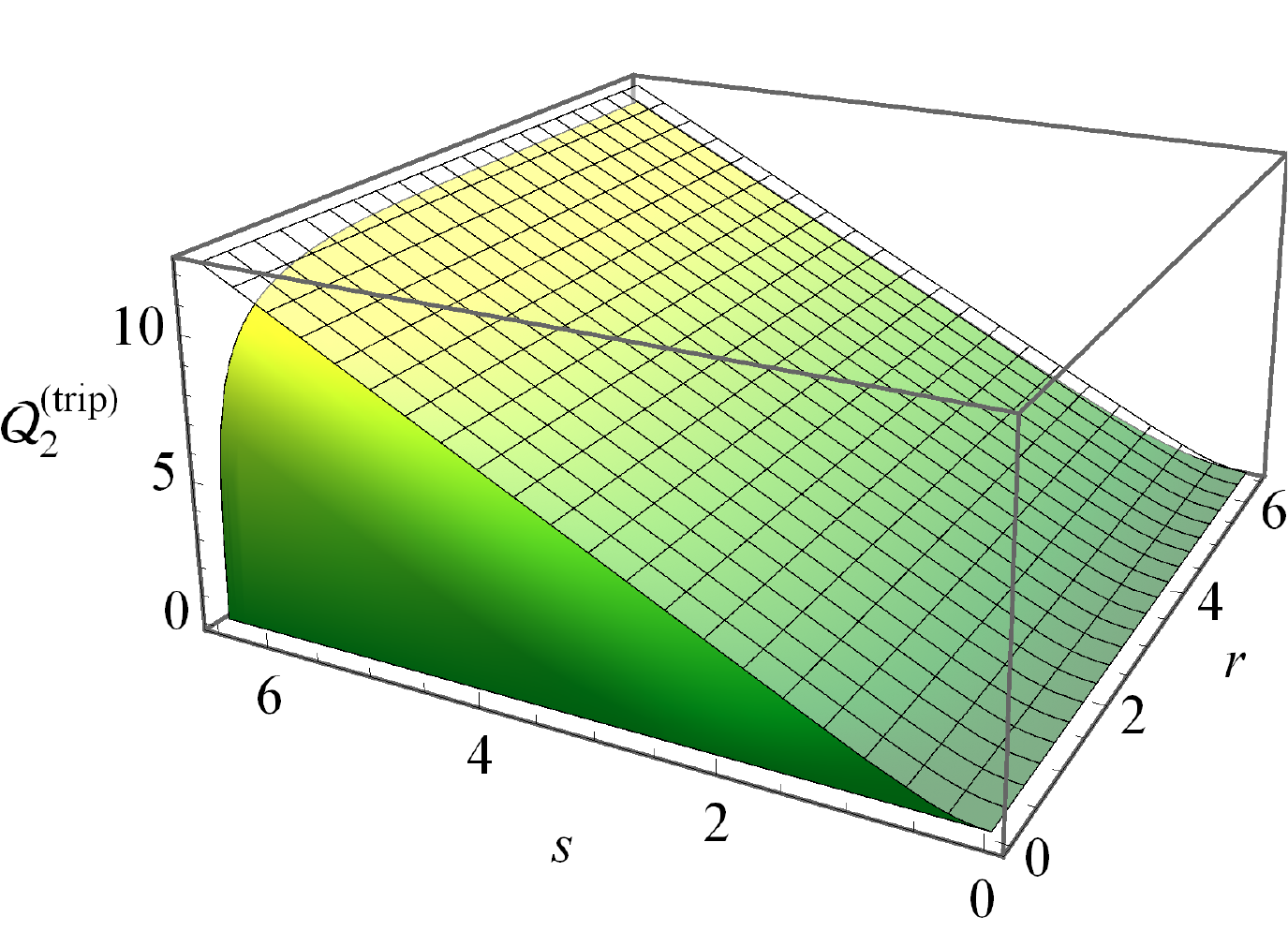}}
\caption{(Color online) Top row: Plots of all the (a) entanglement and (b) discord terms featuring in \eq{arrapaho} as evaluated on the state $\sig_{AR\bar R}^{\rm(a)}$ in the setting in which Alice is inertial while Rob has uniform acceleration parameter $r$. The inertial correlations between modes $A$ and $R$ are characterised by the squeezing parameter $s=\text{arccosh}({\rm e})/2$. The solid green line denotes in both panels the genuine tripartite nonclassical correlations, equally distributed both in the form of residual entanglement and residual discord, and defined as ${\cal Q}_2^{\rm (trip)}(\sig^{\rm(a)}_{A:R:\bar R})$ in \eq{arrapaho}.
Bottom row: Panel (c) depicts the genuine tripartite nonclassical correlations ${\cal Q}_2^{\rm (trip)}(\sig^{\rm(a)}_{A:R:\bar R})$ [shaded surface] as a function of Rob's acceleration parameter $r$ and of the inertial squeezing degree $s$. The correlations ${\cal C}_2\big(\sig^{\rm (M)}_{A:R}\big)$ as described from a fully inertial perspective are depicted as well [wireframe surface].
\label{fivojo}}
\end{center}
\end{figure}

For the sake of the following discussion we can focus on setting (a). It turns out that the genuine tripartite entanglement distributed among $A$, $R$, and $\bar R$ is {\it equal} to the, suitably calculated, genuine tripartite discord. The definition of the genuine tripartite entanglement involves a minimisation over the three possible global bipartitions $A:(R\bar R)$, $R:(A \bar R)$ and $\bar R:(AR)$, as detailed in \cite{ourreview}; similarly, the definition of the genuine tripartite discord involves a minimisation over the mode on which marginal measurements are {\it not} implemented \cite{renyi}. The link between the two is provided by the Koashi-Winter duality relation \cite{koashi}. In the present setting, these minimisations are solved by choosing the $R:(A \bar R)$ splitting and marginal measurements performed on modes other than $R$, respectively. We then obtain, precisely,
\begin{eqnarray}\label{arrapaho}
 &&{\cal E}_2(\sig^{(a)}_{R:(A \bar R)})-{\cal E}_2(\sig^{(a)}_{R:A})-{\cal E}_2(\sig^{(a)}_{R:\bar R}) ={\cal D}_2(\sig^{(a)}_{R|(A \bar R)})-{\cal D}_2(\sig^{(a)}_{R|A})-{\cal D}_2(\sig^{(a)}_{R|\bar R}) \nonumber \\
&\equiv& {\cal Q}_2^{\rm (trip)}(\sig^{\rm(a)}_{A:R:\bar R}) \\
&=& \ln \left[\frac{\left(\cosh ^2(r) \cosh (2 s)+\sinh ^2(r)\right) \left(2 \sinh ^2(r) \cosh (2 s)+\cosh (2 r)+3\right)}{\cosh (2 r) \left((\cosh (2 r)+3) \cosh (2 s)+2 \sinh
   ^2(r)\right)}\right]\,.  \nonumber
\end{eqnarray}
where we have baptised the genuine tripartite nonclassical correlations (merging residual entanglement and residual discord) with the common symbol ${\cal Q}_2^{\rm (trip)}(\sig^{\rm(a)}_{A:R:\bar R})$,
also occasionally referred to as `arravogliament' \cite{contangle}.
We see, interestingly, that although the tripartite entanglement and tripartite discord coincide according to the chosen definitions, the bipartite quantities ${\cal E}_2$ and ${\cal D}_2$ are distributed in a slightly different way across the relevant partitions involving mode $R$, as shown in Fig.~\ref{fivojo}(a) and (b). In particular, as remarked previously, ${\cal E}_2(\sig^{(a)}_{R:A})$ vanishes in the limit of Rob undergoing infinite acceleration $r \rightarrow \infty$ while ${\cal D}_2(\sig^{(a)}_{R|A})$ remains finite [see \eq{dabit}]. As a consequence, we typically get ${\cal D}_2(\sig^{(a)}_{R|\bar R}) \leq {\cal E}_2(\sig^{(a)}_{R:\bar R}$, although both terms increase unboundedly with $r$ (these are the correlations specifically created across the Rindler horizon by the Unruh mechanism).

The genuine tripartite nonclassical correlations ${\cal Q}_2^{\rm (trip)}(\sig^{\rm(a)}_{A:R:\bar R})$ [\eq{arrapaho}] are plotted in Fig.~\ref{fivojo}(c) as a function of $r$ and of the inertial squeezing degree $s$, and compared with the correlations ${\cal C}_2\big(\sig_{A:R}^{\rm (M)}\big)$ [\eq{incorr}] as detectable from a fully inertial perspective. We see that in the limit of high acceleration of Rob and large inertial correlations ($r,s\rightarrow \infty$), a gap of $\ln 2$ remains between the two quantities, meaning that not all inertial correlations are recovered as distributed nonclassical correlations among all involved modes from a noninertial perspective. In fact, the missing bit could be read as the one remaining in the guise of bipartite discord between the two observable modes, \eq{dabit}.

\section{Conclusion and outlook}\label{secconcl}

In this paper we presented a collection of targeted review material and original research, with the primary aim of showcasing the power of continuous variable methods based on Gaussian states and operations, and their natural-born relevance for RQI.
We applied our framework to the now paradigmatic study of degradation of entanglement \cite{entanglementreview} and other forms of quantum and classical correlations \cite{zurek,vedral,modireview} between two scalar field modes in noninertial frames, as a consequence of the Unruh effect \cite{alicefalls,unruh}.

Among the several highlighted phenomena which could be of interest, let us remark that the state of two field modes (with a certain degree of correlations from a fully inertial perspective) as described by an inertial observer Alice and a noninertial observer Rob approaches, in the limit of infinite uniform acceleration of Rob, a so-called classical-quantum state \cite{pianiadesso,modireview}, for which quantum correlations in the form of discord \cite{zurek} are zero if revealed through measurements by Rob, but remain finite if revealed through measurements by Alice, as explained in the previous Section. This could mean that nontrivial quantum communication between Alice and Rob might be possible in one direction only \cite{dattaunruh}.

While these considerations have a value from a foundational perspective, the findings we discussed here have to be regarded more as a teaser, rather than as a concrete setting in which quantum information and communication tasks might be implemented. In our examples, we dealt with the idealised setting of uniformly accelerated observers, and critically with global field modes, whose detection for the purposes of extracting and exploiting correlations as resources in quantum protocols is somewhat troublesome \cite{fay1}.  It is further not obvious how to relate the entanglement and correlation properties of such field modes, as discussed in this paper, to the yield of specific quantum communication settings \cite{bradlercomment}.

Interesting approaches to overcome these theoretical and practical limitations, which are now surfacing in RQI literature, include the application of quantum Shannon theory to characterise the communication capacity associated to relativistic channels \cite{bradler}, the study of entanglement between field modes confined in cavities undergoing general spacetime trajectories \cite{niconew}, the analysis of localised observables as detected in different reference frames for directional quantum communication \cite{nathan}, and novel models of localised field and particle detectors \cite{edu1,edu2}. Crucially, in most of the above settings, {\it mutatis mutandis} one ends up dealing with Gaussian states and transformations. Therefore, the plethora of tools presented here to assess the measure and structure of general types of correlations in bosonic Gaussian states, could and should be readily applied to those more realistic setups, possibly providing new angles for understanding and new pathways for implementation of RQI processing.

This will be the scope of future work.

\section*{Acknowledgments}
We are indebted to Ivette Fuentes who introduced us to the fascinating field of relativistic quantum information, and keeps the local as well as international interest high and active in the area \cite{facebook}. We thank the anonymous referees for highly constructive criticisms and comments on a previous version of this manuscript. GA would like to thank Rob Mann and Tim Ralph for inviting him to write this contribution, and is pleased to acknowledge numerous friends and colleagues with whom he discussed over the years about topics related to this paper, in particular F. Agent, M. Aspachs, D. Bruschi, A. Datta, A. Dragan, M. Ericsson, D. Faccio, N. Friis, I. Fuentes, L. Garay, F. Illuminati, A. Lee, J. Le\'on, J. Louko, R. Mann, E. Mart\'in-Mart\'inez, T. Ralph, A. Serafini.
We thank the University of Nottingham for financial support through an Early Career Research \& Knowledge Transfer Award and a Research Development Fund grant [EPSRC EP/J016314/1 (subcode RDF/BtG/0612b/31)].


\section*{References}
\begin{thebibliography}{99}

\bibitem{peresreview}
Peres A and Terno D R 2004 \rmp {\bf 76}, 93
\bibitem{ternoreview}
Terno D R 2006 {\it Introduction to relativistic quantum information} in {\it Quantum Information Processing: From Theory to Experiment} edited by Angelakis D G {\it et al} (IOP Press) page 61; available as \arx arXiv:quant-ph/0508049
\bibitem{ivettenotes}Fuentes I 2010 {\it Lecture Series on Relativistic Quantum Information} available at\\ \verb"http://iscqi2011.iopb.res.in/talks/revisedLecturenotes_fuentes.pdf"
\bibitem{eduthesis}
Mart\'in-Mart\'inez E 2011 {\it Relativistic Quantum Information: Developments in Quantum Information in general relativistic scenarios} PhD Thesis (Madrid: CSIC); available as \arx arXiv:1106.0280 [quant-ph]
\bibitem{peresterno}
Peres A, Scudo F and Terno D R 2002 \prl {\bf 88} 230402
\bibitem{alicefalls}
Fuentes-Schuller I and Mann R B 2005 \prl {\bf
95}  120404
\bibitem{alicefermion}
Alsing P M, Fuentes-Schuller I, Mann R B and Tessier T E 2006 \pra  {\bf 74} 032326
\bibitem{unruhsharing}
Adesso G, Fuentes-Schuller I and Ericsson M 2007 \pra {\bf 76}
 062112
\bibitem{eduunruh}
Mart\'in-Mart\'inez E, Garay L J and Le\'on J 2010 \prd {\bf 82} 064006;

\bibitem{eduunruh2}
Montero M and Mart\'in-Mart\'inez E 2011 \pra {\bf 84} 012337


\bibitem{bruschi}
Bruschi D E, Louko J, Mart\'in-Mart\'inez E, Dragan A and Fuentes I 2010 \pra {\bf 82} 042332

\bibitem{amplif}
Montero M and Mart\'in-Mart\'inez E 2011 {\it J. High Energ. Phys.} {\bf 07} 006


\bibitem{ball}
Ball J, Fuentes-Schuller I  and Schuller F P 2006 {\it Phys. Lett.} A   {\bf 359} 550

\bibitem{moradi}
Fuentes I, Mann R B, Mart\'in-Mart\'inez E and Moradi S 2010 \prd {\bf 82} 045030

\bibitem{eduparticles}
Mart\'in-Mart\'inez E and Menicucci N C 2012 \arx arXiv:1204.4918 [quant-ph]

\bibitem{retzker}
Reznik B, Retzker A and Silman J 2005 \pra {\bf 71} 042104; Retzker A, Cirac J I and Reznik B 2005 \prl {\bf 94} 050504

\bibitem{ralphtim}
Olson S J and Ralph T C 2011 \prl {\bf 106} 110404; {\it ibid} 2012 \pra {\bf 85}  012306

\bibitem{downes}
Downes T G, Fuentes I and Ralph T C 2011 \prl {\bf 106} 210502

\bibitem{alpha}
Bruschi D E, Fuentes I and Louko J 2012 \prd {\bf 85} 061701(R);
Friis N, Lee A R, Bruschi D E and Louko  J 2012 \prd {\bf 85} 025012

\bibitem{motiongenerates}
Friis N, Bruschi D E, Louko J and Fuentes I 2012 \prd {\bf 85} 081701(R)

\bibitem{niconew}
Friis N and Fuentes I 2012 \arx arXiv:1204.0617 [quant-ph]

\bibitem{alsing}
Alsing P M and Milburn G J 2003 \prl {\bf 91} 180404

\bibitem{bradler}
Br\'adler K, Hayden P and Panangaden P 2009 {\it
Journal of High Energy Physics} {\bf 8} 74;
{\it ibid} 2011 \arx arXiv:1007.0997v3 [quant-ph]

\bibitem{nathan}
Downes T G, Ralph T C and Walk N 2012 \arx
arXiv:1203.2716 [quant-ph]

\bibitem{birelli} Birrell N D and Davies P C W 1982
{\it Quantum fields in curved space} (Cambridge: Cambridge University Press)

\bibitem{nielsenchuang} Nielsen M A and Chuang I L 2000 {\it Quantum Computation and Quantum
  Information} (Cambridge: Cambridge University Press)

\bibitem{eisertplenio}
Eisert J and Plenio M B 2003 {\it Int. J. Quant. Inf.} {\bf 1} 479

\bibitem{brareview}
Braunstein S L and van Loock P 2005 \rmp  {\bf 77} 513

\bibitem{ourreview}
Adesso G and Illuminati F {\bf 2007} {\it J. Phys. A: Math. Theor.} {\bf 40} 7821

\bibitem{book}
Cerf N, Leuchs G and Polzik E S editors 2007  {\it Quantum
Information with
  Continuous Variables of Atoms and Light}\ (London: Imperial College Press)

\bibitem{pirlandolareview}
Weedbrook C {\it et al} 2012 \rmp {\bf 84} 671

\bibitem{ahnkim} Ahn D and Kim M S 2007 \pla {\bf 366} 202

\bibitem{massar} Massar S and Spindel P 2006 \prd
 {\bf 74}  085031

 \bibitem{buchi} Adesso G and Fuentes-Schuller I 2009  {\it Quant. Inf. Comput.} {\bf 9} 0657

\bibitem{mresonance}
Bruschi D E, Dragan A, Lee A R, Fuentes I and Louko J 2012 \arx
 arXiv:1201.0663 [quant-ph]

\bibitem{klauder} Klauder J R and Skagerstam B 1985 {\it Coherent States} (Singapore: World Scientific)

\bibitem{schuch}
 Schuch N, Cirac J I and Wolf M M 2006 {\it  Commun. Math. Phys.} {\bf 267} 65

\bibitem{unruh}
Unruh W G 1976 \prd {\bf 14} 870


\bibitem{hawking}
Hawking S W 1974 {\it Nature} {\bf 248} 30; {\it ibid} 1975 {\it Commun. Math. Phys.} {\bf 43}
199

\bibitem{mariona}
Aspachs M, Adesso G and Fuentes I 2010 \prl {\bf 105} 151301


\bibitem{faccio}  Belgiorno F {\it et al} 2010 {\it Phys. Rev. Lett.} {\bf 105} 203901

\bibitem{com}Schutzhold R and  Unruh W G 2011  {\it Phys. Rev. Lett.} {\bf 107} 149401

\bibitem{fay1}
Dowker F 2011 \arx arXiv:1111.2308 [quant-ph]

\bibitem{edu1}
Dragan A, Doukas J, Martin-Martinez E, and Bruschi D E 2012 \arx arXiv:1203.0655 [quant-ph]

\bibitem{edu2}
Ostapchuk D C M, Lin S Y, Mann R B, and Hu B L 2012  \arx arXiv:1108.3377 [quant-ph]

\bibitem{entanglementreview}
Horodecki R, Horodecki P, Horodecki M and Horodecki K 2009 \rmp {\bf 81} 865

 \bibitem{zurek}
 Ollivier H and Zurek W H 2001 \prl {\bf 88} 017901
 \bibitem{vedral}
 Henderson L and Vedral V 2001 {\it J. Phys. A: Math. Gen.}  {\bf 34} 6899
\bibitem{giordaparis}
Paris M G A and Giorda P 2010 \prl {\bf 105} 020503
 \bibitem{adessodatta}
 Adesso G and Datta A 2010 \prl {\bf 105} 030501
 \bibitem{mistagauss}
  Mi\v{s}ta Jr L, Tatham R, Girolami D, Korolkova N and Adesso G 2011 \pra {\bf  83} 042325
  \bibitem{renyi}
Adesso G, Girolami D and Serafini A 2012 \arx arXiv:1203.5116 [quant-ph]

\bibitem{expdiscordgauss}
Gu M {\it et al} 2012 \arx
 arXiv:1203.0011 [quant-ph]; Blandino R {\it et al} 2012 \arx arXiv:1203.1127 [quant-ph]; Madsen L S, Berni A, Lassen M and Andersen U L 2012
\arx arXiv:1204.2738 [quant-ph]

\bibitem{pianiadesso}
Piani M and Adesso G 2012 \pra  {\bf 85} 040301(R)

 \bibitem{modireview}
 Modi K, Brodutch A, Cable H, Paterek T and Vedral V 2011 \arx arXiv:1112.6238 [quant-ph]

\bibitem{dattaunruh}
Datta A 2009 \pra {\bf 80} 052304

\bibitem{barnett} Walls D F and Milburn G J 1995 {\it Quantum Optics} (Berlin: Springer)

\bibitem{francamentemeneinfischio}
Laurat J {\it et al} 2005  {\it J. Opt. B: Quantum Semiclassical Opt.} {\bf 7} S577

\bibitem{fiurasek07}
 Fiur\'a\v{s}ek J and Mi\v{s}ta Jr L 2007 \pra {\bf 75} 060302(R)

\bibitem{giedkefiurasekdistill}
  Fiur\'a\v{s}ek J 2002 \prl {\bf 89} 137904;
Giedke G and Cirac J I 2002 \pra  {\bf 66} 032316

\bibitem{extremal}
Adesso G, Serafini A and Illuminati F 2004 \pra {\bf 70} 022318

\bibitem{geof}
Wolf M M, Giedke G,  Kr\"uger O, Werner R F and Cirac J I 2004 \pra {\bf 69} 052320; Adesso G and Illuminati F 2005 \pra {\bf 72}
 032334

\bibitem{Arenyi}
R\'{e}nyi A  {\it On measures of information and entropy} 1960 {\it Proc. of the 4th Berkeley Symposium on Mathematics, Statistics and Probability} pages 547--561

\bibitem{holorenyi}
Headrick M 2010 \prd {\bf 82} 126010; Hung J, Myers R C, Smolkin M and Yale A 2011 \arx arXiv:1110.1084v1 [hep-th]

\bibitem{shan}Shannon C E 1948 {\it Bell Syst. Tech. J.} {\bf 27} 623

\bibitem{allegramente}
Giorda P, Allegra M and Paris M G A 2012 \arx
 arXiv:1206.1807 [quant-ph]

\bibitem{unruhreview} Crispino L  {\it et al} 2008 \rmp {\bf 80} 787

\bibitem{squash} Lee A R and Fuentes I 2012 in preparation

\bibitem{contangle}
 Adesso G and Illuminati F 2006 \njp  {\bf 8} 15

\bibitem{koashi}
Koashi M and Winter A 2004 \pra {\bf 69} 022309;



\bibitem{bradlercomment}
Br\'adler K  2011 \arx arXiv:1108.5553v2 [quant-ph]


\bibitem{facebook}
See also the {\it Relativistic Quantum Information} page on Facebook, \\ \verb"https://www.facebook.com/pages/Relativistic-Quantum-Information/207973552600896"
\end{thebibliography}
\end{document}